\documentclass[conference]{IEEEtran}
\IEEEoverridecommandlockouts
\usepackage{cite}
\usepackage{booktabs}
\usepackage{amsmath,amssymb,amsfonts}
\usepackage{algorithm}%
\usepackage{algorithmicx}%
\usepackage{algpseudocode}%
\usepackage{makecell} 
\usepackage{graphicx}
\usepackage{textcomp}
\usepackage{xcolor}
\usepackage{cuted}
\usepackage{stfloats}
\usepackage[caption=false,font=footnotesize]{subfig}
\usepackage{bm} 
\def\BibTeX{{\rm B\kern-.05em{\sc i\kern-.025em b}\kern-.08em
    T\kern-.2em\lower.7ex\hbox{E}\kern-.125emX}}
\usepackage{bm} 
\def\BibTeX{{\rm B\kern-.05em{\sc i\kern-.025em b}\kern-.08em
    T\kern-.2em\lower.7ex\hbox{E}\kern-.125emX}}
\begin{document}

\title{  

Near-Field Dual-UPA Communications: A Generalized Geometric Approach

\thanks{This work is supported in part by the 
Scientific Research Program of Shaanxi Provincial Education Department under Grant 24JK0674 and Natural Science Foundation of Shaanxi Province under Grant 2025JC-YBQN-889.
(\emph{Corresponding author:Xing Hao})
}
 
}
 
\author{     \IEEEauthorblockN{ Li Zheng\textsuperscript{1}, 
    Xing Hao\textsuperscript{2*}, 
    Ziru Chen\textsuperscript{3},
    Yong Liu\textsuperscript{4},
    Li Chen\textsuperscript{1},
    Lin X. Cai\textsuperscript{3}
}

\IEEEauthorblockA{\textsuperscript{1}College of Computer Science, \textsuperscript{2}School of Electronic Information, Northwest University, Xi’an, China}
\IEEEauthorblockA{\textsuperscript{3}Department of Electrical and Computer Engineering, Illinois Institute of Technology, Chicago, USA}
\IEEEauthorblockA{\textsuperscript{4}School of Electronic Science and Engineering, South China Normal University, Foshan, China }

\IEEEauthorblockA{lzheng@stumail.nwu.edu.cn, *xhao@nwu.edu.cn, zchen@ofinno.com, yliu@m.scnu.edu.cn,\\ chenli@nwu.edu.cn, lincai@ieee.org}   
}

\maketitle

\begin{abstract}

This paper investigates a near-field (NF) multiple-input multiple-output (MIMO) communication system equipped with dual uniform planar arrays (UPAs). We first develop a generalized geometric model to calculate the 3D distance between arbitrary antenna elements across the transmitter and receiver panels. Leveraging the distance analysis, we derive a closed-form near-field  to far-field (NF-FF) boundary for dual-UPA configurations. By exploiting the geometric structure of the UPAs, we further decompose the near-field channel matrix into a Kronecker-product of two lower-dimensional matrices. This decomposition enables a low-complexity NF beamforming design for achievable-rate maximization. Numerical results validate the analysis and demonstrate that the conventional Rayleigh distance is a special case of the generalized model. Furthermore, the proposed beamforming design achieves near-optimal rate performance while significantly reducing the computational complexity compared to state-of-the-art NF beamforming methods.

\end{abstract}


\section{Introduction}

6G wireless networks are expected to support high data rates, reliable transmission, and high energy efficiency \cite{6G}. To meet these demands, extremely large-scale multiple-input multiple-output (XL-MIMO) systems \cite{MIMO}, reconfigurable intelligent surfaces (RIS)~\cite{hao2025joint} and high-frequency communication technologies, such as millimeter-wave (mmWave) and terahertz (THz) communications \cite{THz}, have attracted increasing attention. Driven by the enlarged array apertures and shortened carrier wavelengths, the conventional far-field planar-wave model is no longer accurate for characterizing wireless propagation, thereby posing new challenges for beamforming design.

When the propagation distance is much larger than the array aperture, the wireless propagation can be approximated by planar waves, which corresponds to far-field (FF) communication. In this regime, the channel mainly depends on the angular information. As the array aperture increases and the carrier wavelength decreases, the planar-wave assumption becomes inaccurate. In this case, the propagation model should be characterized by spherical-wave \cite{r_nm01}, corresponding to near-field (NF) communication, where the channel depends on both the angle and the distance.

The typical way for  distinguishing the FF and NF communication regions is the classical Rayleigh distance, given by $r_{\mathrm{Rayl}}=2D^2/\lambda$, where $D$ denotes the physical aperture of the antenna array and $\lambda$ is the signal wavelength \cite{Rayleigh}. However, this criterion is mainly based on a single uniform planar array (UPA), i.e., point-to-array propagation. 
With the development of large-scale antenna arrays, dual-UPA MIMO systems, where both the transmitter and receiver are equipped with UPAs, are becoming increasingly relevant for near-field communications. In such systems, the NF-FF boundary may be affected by the joint effect of both array apertures.

As the NF-FF boundary is closely related to the array geometry, geometric analysis has been widely used to characterize spherical-wave propagation and NF-FF region boundaries. Existing studies have modeled spherical-wave propagation using point-to-line configurations for uniform linear arrays (ULAs) \cite{point2line} and point-to-planar configurations for UPAs \cite{point2planar, alamdar202516}, where maximum ratio transmission beamforming is adopted to maximize the received signal-to-noise ratio.  
To the best of our knowledge, the NF-FF boundary analysis and beamforming design for dual-UPA MIMO systems remains underexplored. In addition, due to the high dimensionality of the NF channel matrix in MIMO systems, conventional beamforming designs often require full-dimensional decomposition of the entire channel matrix~\cite{SVD1,SVD2}.
Specifically, the work in \cite{SVD1} maximized the achievable rate by designing the beamforming matrix via the singular value decomposition (SVD) \cite{SVD2} of the entire channel matrix. Such a required full-dimensional SVD may incur high computational complexity, especially for large-scale arrays.

In this paper, we study a near-field MIMO wireless communication network, where both the transmitter and the receiver are equipped with UPAs. We first develop a generalized geometric model to calculate the 3D distance between arbitrary antenna elements across the transmitter and receiver panels. The 3D distance is mathematically proven to be formulated as two independent additive terms. Based on this model, we derive a closed-form boundary between the NF and FF regions as a function of the wavelength and the apertures of the transmitter and receiver UPAs. Furthermore, by exploiting the geometric structure of the UPAs, the NF channel matrix is expressed in a Kronecker-product form with two lower-dimensional matrices. We then develop a closed-form, low-complexity beamforming design for achievable-rate maximization. Finally, numerical results validate the derived boundary and show that the conventional Rayleigh distance is a special case of the proposed boundary, while the proposed beamforming design achieves near-optimal rate performance with reduced computational complexity.

\section{system model and problem formulation}

\subsection{Array Geometry and Coordinate System}

\begin{figure}[t]
\centerline{\includegraphics[width=0.4\textwidth]{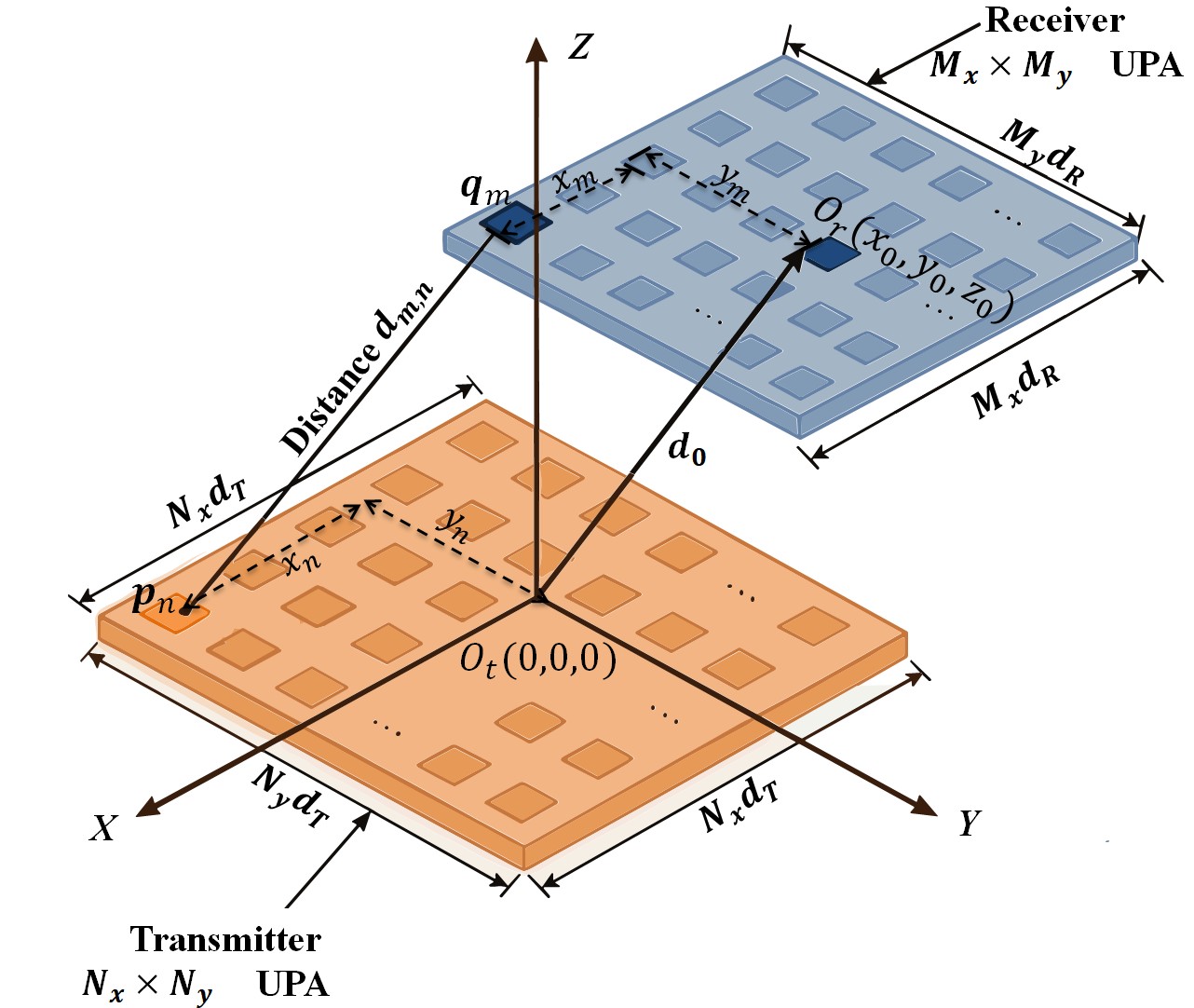}}
\caption{ Geometric Model of the Near-Field MIMO wireless communications with a 3D Coordinate System}
\label{DualMIMO}
\end{figure}

As shown in Fig.  \ref{DualMIMO}, we consider a downlink near-field MIMO communication network, where  the transmitter and the receiver are both equipped with UPAs.  
To characterize the array geometry, we establish a three-dimensional (3D) Cartesian coordinate system.  
The geometric center of the transmitter UPA is taken as the origin $O_t(0,0,0)$, and the array is assumed to be located on the $x-y$ plane (i.e., $z=0$). The transmitter UPA comprises $N = N_x N_y$ antennas, where $N_x$ and $N_y$ denote the numbers of antennas along the $x$-axis and $y$-axis, respectively. The spacing between adjacent transmit antennas is denoted by $d_T$ and the aperture of the transmitter UPA is $D_T=\sqrt{(N_x d_T)^2 +(N_y d_T)^2}$.
Let  $O_r(x_0, y_0, z_0)$ denote the geometric center of the receiver UPA, and the  center-to-center distance between the transmitter UPA and the receiver UPA is defined as  
\begin{equation}
d_0=\sqrt{x_0^2 +  y_0^2 + z_0^2} .
\label{eq:d_0}
\end{equation}
The receiver UPA is parallel to the transmitter UPA and consists of $M = M_x M_y$ antennas, where $M_x$ and $M_y$ denote the numbers of antennas aligned with the $x$-axis and $y$-axis, respectively.  The spacing between adjacent receive antennas is denoted by $d_R$ and the aperture of the receiver UPA is $D_R=\sqrt{(M_x d_R)^2 +(N_y d_R)^2}$.

Without loss of generality, let $\mathbf{p}_n$ and $\mathbf{q}_m$ denote the position vectors of an arbitrary antenna on the transmitter UPA and the receiver UPA, respectively.  Let $( x_n, y_n)$ denote the relative planar coordinates of the considered transmitter antenna with respect to the transmitter UPA center,  and let $(x_m, y_m)$ denote the planar coordinates of the considered receiver antenna with respect to the center of the receiver UPA. Since the arrays are parallel to the $x-y$ plane, their position vectors in the coordinate system are given by
\begin{equation}
\mathbf{p}_n=[x_n, y_n,0],\quad \mathbf{q}_m=[x_0+x_m, y_0+y_m, z_0].
\end{equation}
Accordingly, the Euclidean distance between the considered transmit antenna and receive antenna is given by
\begin{equation}
d_{m,n} =  \sqrt{(x_0+x_m - x_n)^2 + ( y_0+y_m - y_n)^2 + z_0^2}.
\label{eq:distance}
\end{equation}

\subsection{ Communication Model}

Under the uniform spherical wave model, the near-field channel coefficient $h_{m,n}$ between the considered transmit and receive antennas is directly determined by the derived distance $d_{m,n}$ in \eqref{eq:distance}. Specifically, it is given by

\begin{equation}
h_{m,n}=\frac{\lambda}{4\pi d_{0}}e^{-j\frac{2\pi}{\lambda}d_{m,n}},
\label{eq:channel_element2}
\end{equation}
where $\lambda$ is the carrier wavelength. 
Let $\mathbf{H} \in \mathbb{C}^{M \times N} $ denotes the near-field channel matrix between the transmitter and the receiver.  The channel matrix $\mathbf{H}$  is formed by the channel coefficients in \eqref{eq:channel_element2}, given by
\begin{equation}
\mathbf{H} = \begin{bmatrix}
h_{1,1} & h_{1,2} & \cdots & h_{1,N} \\
h_{2,1} & h_{2,2} & \cdots & h_{2,N} \\
\vdots & \vdots & \ddots & \vdots \\
h_{M,1} & h_{M,2} & \cdots & h_{M,N} \end{bmatrix}, 
\label{channel}
\end{equation}

Let $\mathbf{s}\in\mathbb{C}^{L}$ denote the transmitted data stream vector, where $L\leq \min\{M,N\}$ is the number of data streams. Let $\mathbf{W}\in\mathbb{C}^{N\times L}$ denote the transmit beamforming matrix at the transmitter. The transmitted signal is given by
$\mathbf{x}=\mathbf{W}\mathbf{s}$. During the downlink transmission, the received signal  at the receiver is expressed as
\begin{equation}
\mathbf{y}=\mathbf{H}\mathbf{W}\mathbf{s}+\mathbf{n},
\label{eq:received_signal}
\end{equation}
where $\mathbf{n}\sim\mathcal{CN}(\mathbf{0},\sigma^2\mathbf{I}_M)$ is the additive white Gaussian noise vector.

\subsection{Problem Formulation}

According to the received signal model in \eqref{eq:received_signal}, the achievable rate of the receiver is given by
\begin{equation}
R=\log_2\det\!\left(\mathbf{I}_M+\frac{1}{\sigma^2}\mathbf{H}\mathbf{W}\mathbf{W}^H\mathbf{H}^H\right).
\end{equation}

The transmit beamforming design is formulated as the following achievable-rate maximization problem under the transmit power constraint
\begin{subequations}
\begin{align}
(\mathcal{P}_1):\quad \max_{\mathbf{W}} \quad & R \\
\text{s.t.}\quad & \mathrm{Tr}(\mathbf{W}\mathbf{W}^H)\le P_{\max}.
\end{align}
\end{subequations}
where $P_{\max}$ denotes the maximum transmit power at the transmitter.

\section{Boundary Between Near-field and Far-field Regions}

We first process the exact Euclidean distance $d_{m,n}$ derived in the system model. By expanding the quadratic terms inside the square root and extracting the center-to-center distance $d_0$,  the \eqref{eq:distance}  is rewritten as  as \eqref{eq:exact_expanded}, shown at the bottom of this page.
\begin{figure*}[!b]
\hrulefill
\begin{equation}
\begin{aligned}
d_{m,n} &= \sqrt{d_0^2 + (x_m - x_n)^2 + 2x_0(x_m - x_n) + (y_m - y_n)^2 + 2y_0(y_m - y_n)}\\
&= d_0 \sqrt{1 + \frac{(x_m - x_n)^2 + 2x_0(x_m - x_n) + (y_m - y_n)^2 + 2y_0(y_m - y_n)}{d_0^2}}\\
\end{aligned}
\label{eq:exact_expanded}
\end{equation}
\end{figure*}
Since the array aperture is typically much smaller than the transmission distance, the fractional term inside the square root is significantly less than 1. By applying the Taylor expansion \cite{Taylor1,Taylor2}, i.e., $\sqrt{1+x} \approx 1 + \frac{x}{2}$ for $|x|<1$, the distance is given by

\begin{equation}
\begin{aligned}
d_{m,n} &\approx 
 d_0 + \frac{x_m^2 \!+\! x_n^2 \!+\! 2x_0 x_m \!-\! 2x_0 x_n \!-\! 2x_m x_n}{2d_0} \\
&\quad + \frac{y_m^2 \!+\! y_n^2 \!+\! 2y_0 y_m \!-\! 2y_0 y_n \!-\! 2y_m y_n}{2d_0}.\\
\end{aligned}
\label{eq:taylor_expansion}
\end{equation}

To simplify notation and facilitate the matrix decomposition, let $d_{m,n}^x$ and $d_{m,n}^y$ be algebraic components decoupled from the horizontal and vertical coordinates, respectively, given by

\begin{equation}
\begin{aligned}
  d_{m,n}^x &= \frac{x_m^2 + x_n^2 + 2x_0 x_m - 2x_0 x_n - 2x_m x_n}{2d_0},
\end{aligned} 
\end{equation}

\begin{equation}
\begin{aligned}
  d_{m,n}^y &= \frac{y_m^2 + y_n^2 + 2y_0 y_m - 2y_0 y_n - 2y_m y_n}{2d_0}.
\end{aligned} 
\end{equation}
The  \eqref{eq:taylor_expansion} can be rewritten as  $d_{m,n} =d_0 + d_{m,n}^x + d_{m,n}^y$. The corresponding phase shift cause by the signal propagation through $d_{m,n}$ is $\phi_{m,n}=-\frac{2\pi d_{m,n}}{\lambda} $.

If we use the phase shift caused by the wireless propagation between the centers of the transmitter and receiver UPAs as the reference, denoted by $\phi_{{ref}}$, the  relative phase difference associated with the channel between the $n$-th transmit antenna and the $m$-th receive antenna, denoted by $\hat{\phi}_{m,n}$, can be formulated as

\begin{equation}
\hat{\phi}_{m,n} = \phi_{m,n} - \phi_{{ref}} = -\frac{2\pi}{\lambda}(d_{m,n} - d_0).
\end{equation}

Under the planar-wave assumption, the far-field propagation distance  between the $n$-th transmit antenna and the $m$-th receive antenna is written as
\begin{equation}
\begin{aligned}
d_{m,n}^{\rm far}=d_0+\frac{x_0(x_m-x_n)+y_0(y_m-y_n)}{d_0}.
\end{aligned}
\label{eq:far_distance}
\end{equation}
Accordingly, the distance difference between $d_0$ and $d_{m,n}^{\rm far}$ is given by $d_0-d_{m,n}^{\rm far} =-\frac{x_0(x_m-x_n)+y_0(y_m-y_n)}{d_0}$.
The corresponding relative phase difference under the planar-wave assumption, denoted by $\hat{\phi}_{m,n}^{ far}$, can be written as

\begin{equation}
\begin{aligned}
\hat{\phi}_{m,n}^{ far}
&=\phi_{m,n}^{ far}-\phi_{{ref}} 
= -\frac{2\pi}{\lambda}  \frac{x_0(x_m-x_n)+y_0(y_m-y_n)}{d_0}.
\end{aligned}
\label{eq:far_phase_difference}
\end{equation}

Let $\Delta \Phi_{m,n}$ denote the phase difference for the link between the $n$-th transmit antenna and the $m$-th receive antenna. Based on \eqref{eq:taylor_expansion} and \eqref{eq:far_distance},   $\Delta \Phi_{m,n}$ is given by
\begin{equation}
\begin{aligned}
    \Delta \Phi_{m,n} &= \left| \hat{\phi}_{m,n} - \hat{\phi}_{m,n}^{\text{far}} \right|\\
    &= \frac{2\pi}{\lambda} \left| d_{m,n} - d_{m,n}^{\text{far}} \right|\\
    &= \frac{\pi}{\lambda d_0}  [{(x_m-x_n)^2 + (y_m-y_n)^2}].
    \label{PD}
\end{aligned}
\end{equation}

The boundary between the near-field and far-field regions is defined as the boundary where the maximum phase difference of the received signal by using spherical wave and planar wave assumption on the antenna array does not exceed a given threshold $\overline{\phi}$. 
Therefore, the planar-wave approximation is inaccurate if the maximum phase difference satisfy the threshold constraint

\begin{equation}
\max_{m,n} \frac{2\pi}{\lambda} \frac{(x_m-x_n)^2 + (y_m-y_n)^2}{2d_0} \le \overline{\phi}.
\end{equation}
This condition leads to the following boundary

\begin{equation}
d_0 \ge    \max_{m,n}  \frac{\pi \left(   (x_m-x_n)^2 + (y_m-y_n)^2 \right)}{\lambda \overline{\phi}}.
\label{d_Rayleigh1}
\end{equation}

 Since the  coordinates of the transmit and receive antennas are defined with respect 
to their array centers, we have $\sqrt{x_n^2+y_n^2}\leq \frac{D_T}{2}$ and $\sqrt{x_m^2+y_m^2}\leq \frac{D_R}{2}$.
Based on the triangle inequality, we have 
\begin{equation}
\begin{aligned}
\sqrt{(x_m-x_n)^2+(y_m-y_n)^2} &\leq  \sqrt{x_m^2+y_m^2} + \sqrt{x_n^2+y_n^2}\\&  \leq \frac{D_T+D_R}{2}.
\end{aligned}
\end{equation}
It follows that
\begin{equation}
\begin{aligned}
(x_m-x_n)^2+(y_m-y_n)^2 \leq \frac{(D_T+D_R)^2}{4}.
\end{aligned}
\end{equation}
Accordingly, the boundary \eqref{d_Rayleigh1} can be sufficiently guaranteed by
\begin{equation}
d_0
\geq
\frac{\pi(D_T+D_R)^2}{4\lambda\bar{\phi}}.
\label{eq:rayleigh_DT_DR}
\end{equation}

\section{Beamforming Design}

In this section, we propose a geometry-based low-complexity beamforming design for problem ($\mathcal{P}_1$). Specifically, based on the distance reformulation in \eqref{eq:taylor_expansion},  we first express the NF channel matrix in a Kronecker-product form.  A Kronecker-SVD-based beamforming design is then developed by performing two low-dimensional SVDs followed by water-filling power allocation. Finally, the computational complexity of the proposed design is analyzed and compared with that of the conventional full-dimensional SVD-based method.

\subsection{Kronecker Product-Based Channel}

Without loss of generality, the channel $h_{m,n}$ between the $m$-th receive antenna and the $n$-th transmit antenna  can be rewritten as

\begin{equation}
\begin{aligned}
h_{m,n} &=\frac{\lambda}{4\pi d_0} e^{-j\frac{2\pi}{\lambda}  d_{m,n} } \\
&\approx \frac{\lambda}{4\pi d_0} e^{-j\frac{2\pi}{\lambda}(d_0 + d_{m,n}^x + d_{m,n}^y)} \\
&= \gamma_0 e^{-j\frac{2\pi}{\lambda}d_{m,n}^x} e^{-j\frac{2\pi}{\lambda}d_{m,n}^y},
\end{aligned}
\label{eq:scalar_factorization}
\end{equation}
where $\gamma_0 = \frac{\lambda}{4\pi d_0} e^{-j\frac{2\pi}{\lambda}d_0}$ denotes the  complex channel attenuation and phase shift determined by the center-to-center distance $d_0$. The \eqref{eq:scalar_factorization} shows that each channel coefficient is factorized into  a $\gamma_0 $, a term depending only on the $y$-coordinates, and a term depending only on the $x$-coordinates.

According to the algebraic definition of the Kronecker product, this element-wise multiplication  translates into a matrix-level product of two lower-dimensional exponential matrices. Specifically, the  channel matrix $\mathbf{H}$ can be written as

\begin{equation}
\begin{aligned}
\mathbf{H}
= \gamma_0 &\begin{bmatrix} e^{-j\frac{2\pi}{\lambda}(d_{1,1}^x + d_{1,1}^y)} & \cdots & e^{-j\frac{2\pi}{\lambda}(d_{1,N}^x + d_{1,N}^y)} \\ \vdots & \ddots & \vdots \\ e^{-j\frac{2\pi}{\lambda}(d_{M,1}^x + d_{M,1}^y)} & \cdots & e^{-j\frac{2\pi}{\lambda}(d_{M,N}^x + d_{M,N}^y)} \end{bmatrix}\\
= \gamma_0   &\begin{bmatrix} e^{-j\frac{2\pi}{\lambda}d_{1,1}^y} & \cdots & e^{-j\frac{2\pi}{\lambda}d_{1,N_y}^y} \\ \vdots & \ddots & \vdots \\ e^{-j\frac{2\pi}{\lambda}d_{M_y,1}^y} & \cdots & e^{-j\frac{2\pi}{\lambda}d_{M_y,N_y}^y} \end{bmatrix}   \\ 
 \quad   &\otimes \begin{bmatrix} e^{-j\frac{2\pi}{\lambda}d_{1,1}^x} & \cdots & e^{-j\frac{2\pi}{\lambda}d_{1,N_x}^x} \\ \vdots & \ddots & \vdots \\ e^{-j\frac{2\pi}{\lambda}d_{M_x,1}^x} & \cdots & e^{-j\frac{2\pi}{\lambda}d_{M_x,N_x}^x} \end{bmatrix}  \\
= \gamma_0 &(\mathbf{H}_y \otimes \mathbf{H}_x)  
\label{eq:Kronecker_H}
\end{aligned}
\end{equation}
where $\otimes$ denotes the Kronecker product,  $\mathbf{H}_y \in \mathbb{C}^{M_y \times N_y}$ and $\mathbf{H}_x \in \mathbb{C}^{M_x \times N_x}$ denote the low-dimensional component matrices associated with the vertical and horizontal dimensions, respectively. Specifically, the entries in the $m_y$-th row, $n_y$-th column of $\mathbf{H}_y$ and the $m_x$-th row, $n_x$-th column of $\mathbf{H}_x$ are  given by
\begin{equation}
[\mathbf{H}_y]{m_y,n_y} = e^{-j\frac{2\pi}{\lambda}d_{m,n}^y}, \ [\mathbf{H}_x]{m_x,n_x} = e^{-j\frac{2\pi}{\lambda}d_{m,n}^x}.
\end{equation}
These component matrices are   mathematical constructs derived from the algebraic distance decoupling, which provide a tractable Kronecker product structure to reduce the computational complexity of beamforming design.

\subsection{Near Field Beamforming Design}

For the achievable rate maximization problem in $(\mathcal{P}_1)$, we exploit the Kronecker product structure of the channel matrix derived in \eqref{eq:Kronecker_H} to propose a low-complexity beamforming design. Specifically,   instead of performing the SVD on the  channel matrix $\mathbf{H}$, we apply the SVD independently on the low-dimensional  component matrices in~\eqref{eq:Kronecker_H}.  We  perform the SVDs of $\mathbf{H}_x$ and $\mathbf{H}_y$ as

\begin{equation}
\mathbf{H}_x = \mathbf{U}_x \mathbf{\Sigma}_x \mathbf{V}_x^H, \quad \mathbf{H}_y = \mathbf{U}_y \mathbf{\Sigma}_y \mathbf{V}_y^H,
\label{SVDs}
\end{equation}
where $\mathbf{U}_x \in \mathbb{C}^{M_x \times M_x}$, $\mathbf{V}_x \in \mathbb{C}^{N_x \times N_x}$, $\mathbf{U}_y \in \mathbb{C}^{M_y \times M_y}$, and $\mathbf{V}_y \in \mathbb{C}^{N_y \times N_y}$ are unitary matrices. The matrices $\mathbf{\Sigma}_x \in \mathbb{C}^{M_x \times N_x}$ and $\mathbf{\Sigma}_y \in \mathbb{C}^{M_y \times N_y}$ are rectangular diagonal matrices containing the singular values of the respective dimensions.  

By substituting \eqref{SVDs}   into \eqref{eq:Kronecker_H} and employing the mixed-product property of the Kronecker product, i.e., $(\mathbf{A} \otimes \mathbf{B})(\mathbf{C} \otimes \mathbf{D}) = (\mathbf{A}\mathbf{C}) \otimes (\mathbf{B}\mathbf{D})$,  $\mathbf{H}$ is reformulated  as

\begin{equation}
\begin{aligned}
\mathbf{H} &= \gamma_0 \left( \mathbf{U}_y \mathbf{\Sigma}_y \mathbf{V}_y^H \right) \otimes \left( \mathbf{U}_x \mathbf{\Sigma}_x \mathbf{V}_x^H \right) \\
&= \gamma_0 (\mathbf{U}_y\otimes \mathbf{U}_x) (\mathbf{\Sigma}_y\otimes \mathbf{\Sigma}_x) (\mathbf{V}_y\otimes \mathbf{V}_x)^H\\
&= \mathbf{U}_{\text{eq}} \mathbf{\Sigma}_{\text{eq}} \mathbf{V}_{\text{eq}}^H,
\end{aligned}
\end{equation}
where $\mathbf{U}_{\text{eq}} = \mathbf{U}_y \otimes \mathbf{U}_x \in \mathbb{C}^{M \times M}$ and $\mathbf{V}_{\text{eq}} = \mathbf{V}_y \otimes \mathbf{V}_x \in \mathbb{C}^{N \times N}$ serve as the left and right singular matrices, respectively. The equivalent singular value matrix is given by $\mathbf{\Sigma}_{\text{eq}} = \gamma_0 (\mathbf{\Sigma}_y \otimes \mathbf{\Sigma}_x) \in \mathbb{C}^{M \times N}$, and  the singular values in $\mathbf{\Sigma}_{\text{eq}}$ can be given by the pairwise products
\begin{equation}
\lambda_{i,j} = |\gamma_0| \, \sigma_{y,i} \sigma_{x,j}, \quad i=1,\dots,r_y,\ j=1,\dots,r_x,
\end{equation}
where $r_x = \min(M_x, N_x)$ and $r_y = \min(M_y, N_y)$ denote the maximum numbers of non-zero singular values for the horizontal and vertical dimensions, respectively. 
$\sigma_{x,j}$ and $\sigma_{y,i}$ denote the diagonal entries of $\mathbf{\Sigma}_x$ and $\mathbf{\Sigma}_y$, respectively. We sort $\{\lambda_{i,j}\}$ in descending order and select the $L$ largest values. Let $\widetilde{\mathbf{V}}_L\in\mathbb{C}^{N\times L}$ denote the matrix formed by the corresponding $L$ columns of $\mathbf{V}_y\otimes \mathbf{V}_x$.
By applying the determinant identity $\det(\mathbf{I} + \mathbf{AB}) = \det(\mathbf{I} + \mathbf{BA})$ and  the unitary invariance property $\det(\mathbf{I} +\mathbf{U} \mathbf{A}\mathbf{U} ^H) = \det(\mathbf{I} + \mathbf{A})$, the proposed beamforming matrix is given  by

\begin{equation}
\mathbf{W}_{\mathrm{prop}} =\widetilde{\mathbf{V}}_L\mathbf{P}^{1/2},
\label{eq:W_prop}
\end{equation}
where $\mathbf{P}=\mathrm{diag}\{p_1,\dots,p_d\}\in\mathbb{C}^{L\times L}$ is the power allocation matrix, and the optimal power $p_\ell$ is obtained by the water-filling algorithm \cite{WF} under the transmit power constraint $\sum_{\ell=1}^{L} p_\ell \le P_{\max}$.

\subsection{Complexity Analysis}

We compare the computational complexity of the conventional SVD beamforming design and the proposed Kronecker-SVD-based design.
For the conventional method, the main computational burden comes from performing the  SVD of the  channel matrix, i.e., $\mathbf{H}=\mathbf{U}\mathbf{\Sigma}\mathbf{V}^H$, where $\mathbf{U} \in \mathbb{C}^{M \times M}$ and $\mathbf{V} \in \mathbb{C}^{N \times N}$ are unitary matrices, and $\mathbf{\Sigma} \in \mathbb{C}^{M \times N}$ is a rectangular diagonal matrix. The corresponding complexity is 
\begin{equation}
\mathcal{O}\!\left(M_xM_yN_xN_y\min(M_xM_y,N_xN_y)\right).
\label{eq:complexity_full}
\end{equation}

For the proposed method,  we only need to compute the SVDs of the two lower-dimensional channel matrices $\mathbf{H}_x\in\mathbb{C}^{M_x\times N_x}$ and $\mathbf{H}_y\in\mathbb{C}^{M_y\times N_y}$. Therefore, the complexity of the proposed Kronecker-SVD method is
\begin{equation}
\mathcal{O}\!\left(M_xN_x\min(M_x,N_x)+M_yN_y\min(M_y,N_y)\right).
\label{eq:complexity_prop_svd}
\end{equation}

\section{Simulation Results}

In this section, we present numerical results to evaluate the proposed near-far-field boundary analysis and the low-complexity beamforming design. We first validate the derived boundary by showing the boundary distance under different numbers of transmitter antennas, phase difference thresholds, and carrier frequencies. The results also illustrate the relationship between the proposed boundary and the conventional Rayleigh distance. Then, we evaluate the proposed Kronecker-SVD beamforming design in terms of achievable rate and CPU running time. For comparison, we consider several benchmark schemes, including full-dimensional SVD over the NF channel with water-filling, full-dimensional SVD over the NF channel with equal power allocation, and far-field SVD beamforming based on the planar-wave channel approximation.

\subsection{Simulation Setup}

We consider a downlink near-field planar MIMO system operating at a carrier frequency of $f_c=60$ GHz, with the corresponding wavelength denoted by $\lambda=0.005 $ m. Both the transmitter  and the receiver are equipped with UPAs,  where the antenna spacing is set to a half-wavelength, i.e., $d_T = d_R = \lambda/2$. The transmitter  is located at the origin of the 3D coordinate system, and the receiver UPA is placed parallel to the transmitter  UPA at a center-to-center distance of $r_0$, where $r_0$ is assumed to be within the NF region. The number of transmitter  antennas is set to $N = N_x \times N_y = 40 \times 40$, and the number of receiver antennas is $M = M_x \times M_y = 16 \times 16$. The number of transmitted data streams is set to  $L=4\leq \min\{M,N\}$ and the maximum transmit power at the transmitter  is set to $P_{max} = 10$ dBm.

\subsection{Numerical Results}

\begin{figure}
\centerline{\includegraphics[width=0.45\textwidth]{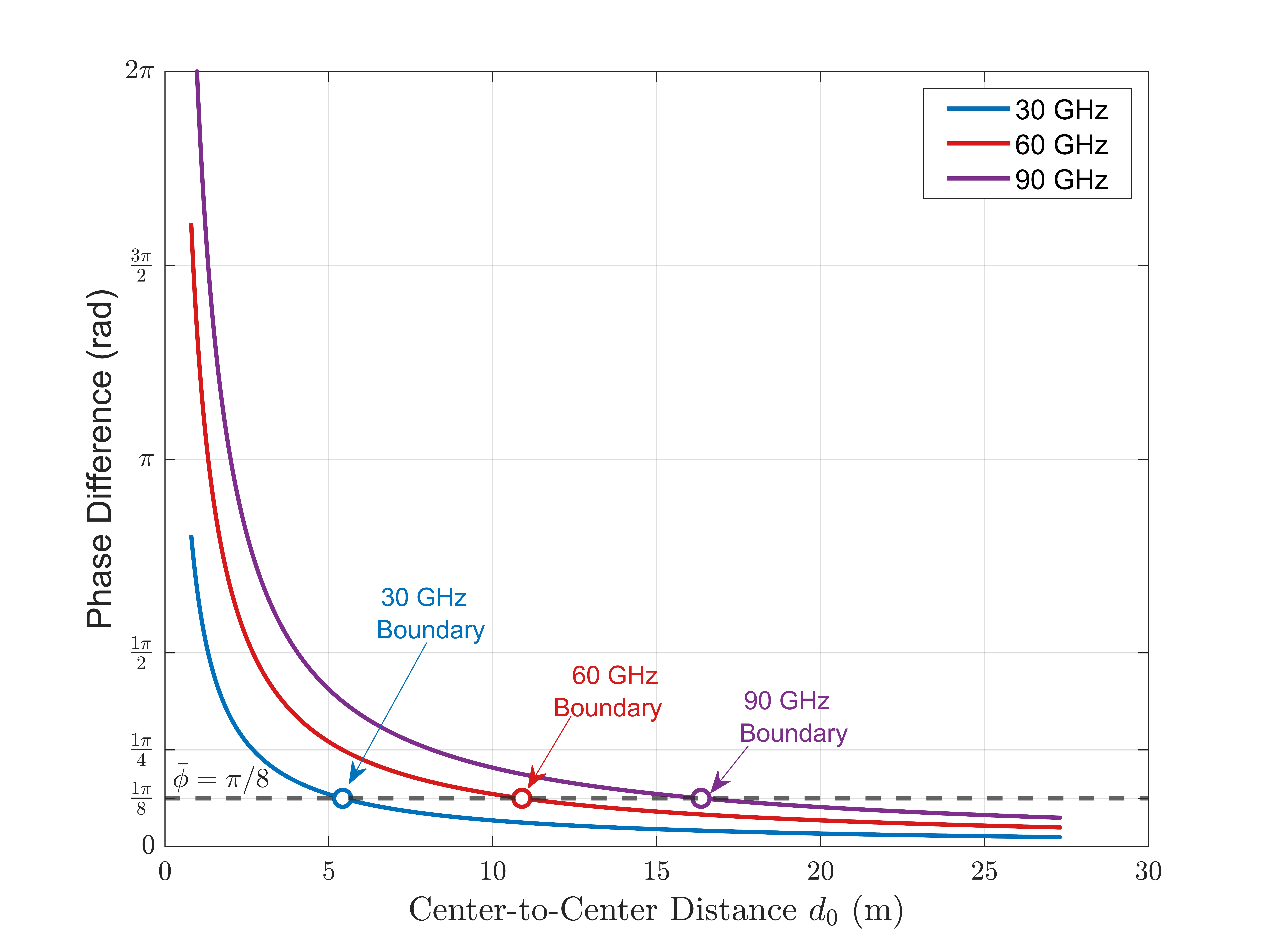}}
\caption{The phase difference  versus center-to-center distance.}
\label{GHZ}
\end{figure}

Fig. \ref{GHZ} shows the phase difference $\Delta \Phi_{m,n}$ between the near-field spherical-wave model and the far-field planar-wave model versus the center-to-center distance under different carrier frequencies. For the same propagation distance, a higher carrier frequency results in a larger phase difference, since the wavelength $ \lambda$ becomes shorter and the phase difference is proportional to $1/ \lambda$. Therefore, under the same  threshold  $\bar{\phi}=\pi/8$, the derived boundary distance becomes larger as the carrier frequency increases. where  Specifically, the 90 GHz case requires the largest boundary distance, while the 30 GHz case has the smallest one. This result shows that high-frequency systems are more likely to operate in the near-field region, which further supports the necessity of spherical-wave channel modeling for mmWave/THz MIMO systems.

\begin{figure}
\centerline{\includegraphics[width=0.45\textwidth]{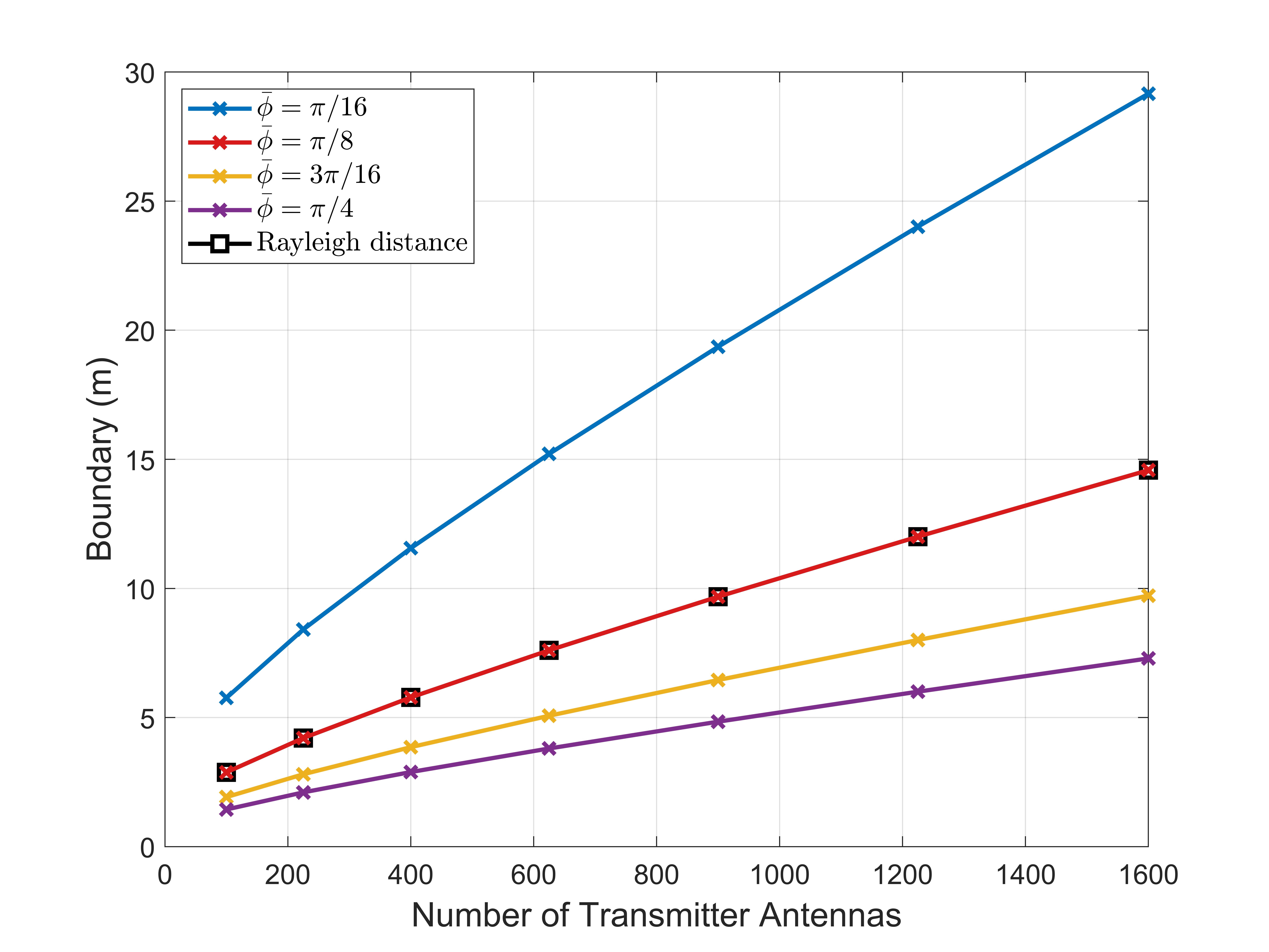}}
\caption{ NF-FF boudnary in~\eqref{eq:rayleigh_DT_DR} versus the number of transmitter antenna.}
\label{VsRayleigh}
\end{figure}
Fig. \ref{VsRayleigh} illustrates the boundary distance calculated from \eqref{eq:rayleigh_DT_DR} versus the number of transmitter antennas under different threshold $\overline{\phi}$. It can be observed that the boundary distance increases with the number of transmitter antennas. This is because increasing the number of antennas enlarges the transmitter aperture. According to \eqref{eq:rayleigh_DT_DR}, the derived boundary distance scales with $(D_T+D_R)^2$ and thus increases with the array aperture.  It is also observed that the proposed boundary is equivalent to the Rayleigh distance when $\bar{\phi}=\pi/8$. Moreover, a smaller threshold leads to a larger boundary distance, which is consistent with the inverse proportional relationship between the boundary distance and $\bar{\phi}$.

\begin{figure}
\centerline{\includegraphics[width=0.45\textwidth]{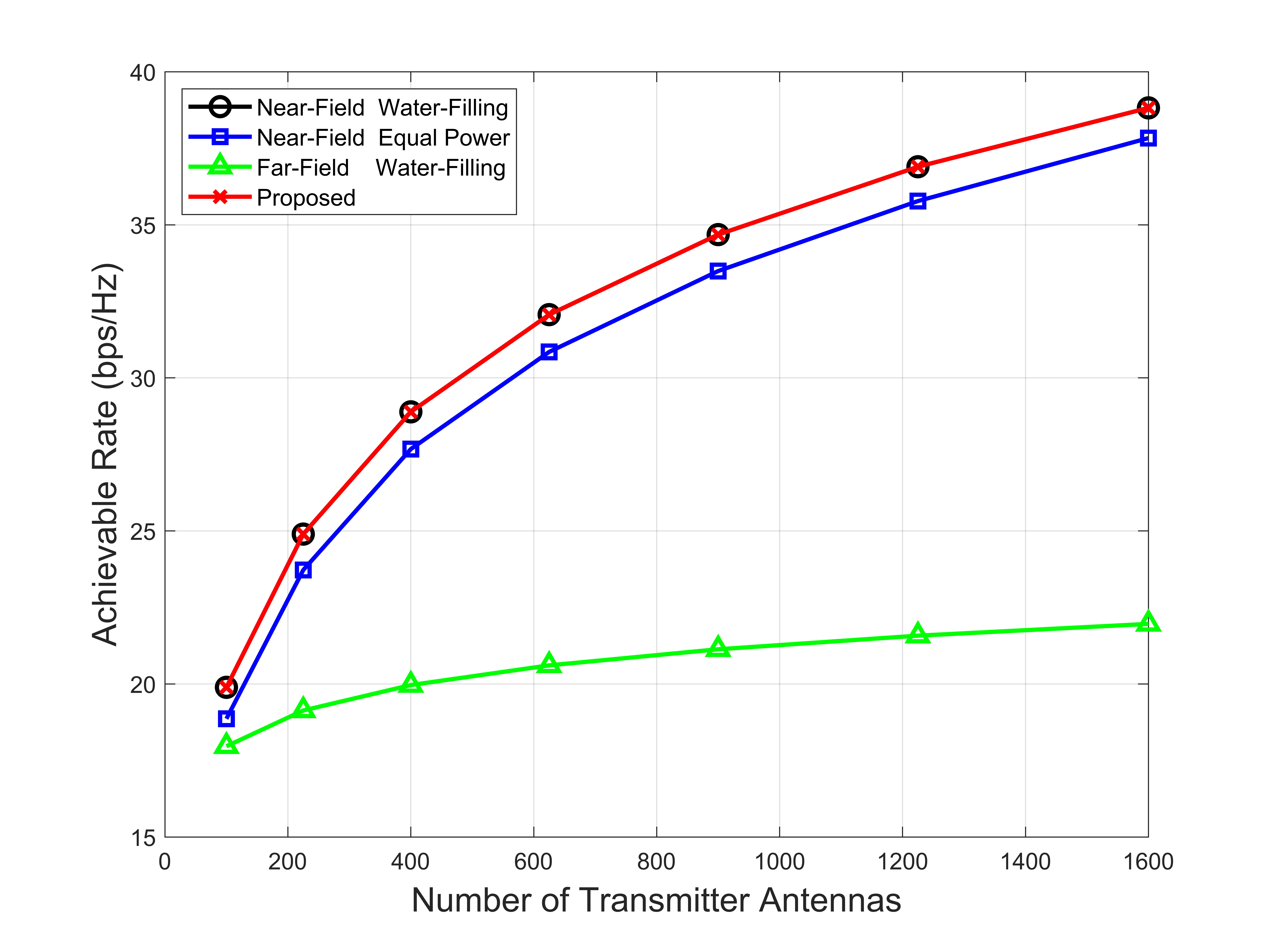}}
\caption{The achievable rate  versus the number of transmitter  antenna.}
\label{transmitter number_rate}
\end{figure}

\begin{figure}
\centerline{\includegraphics[width=0.45\textwidth]{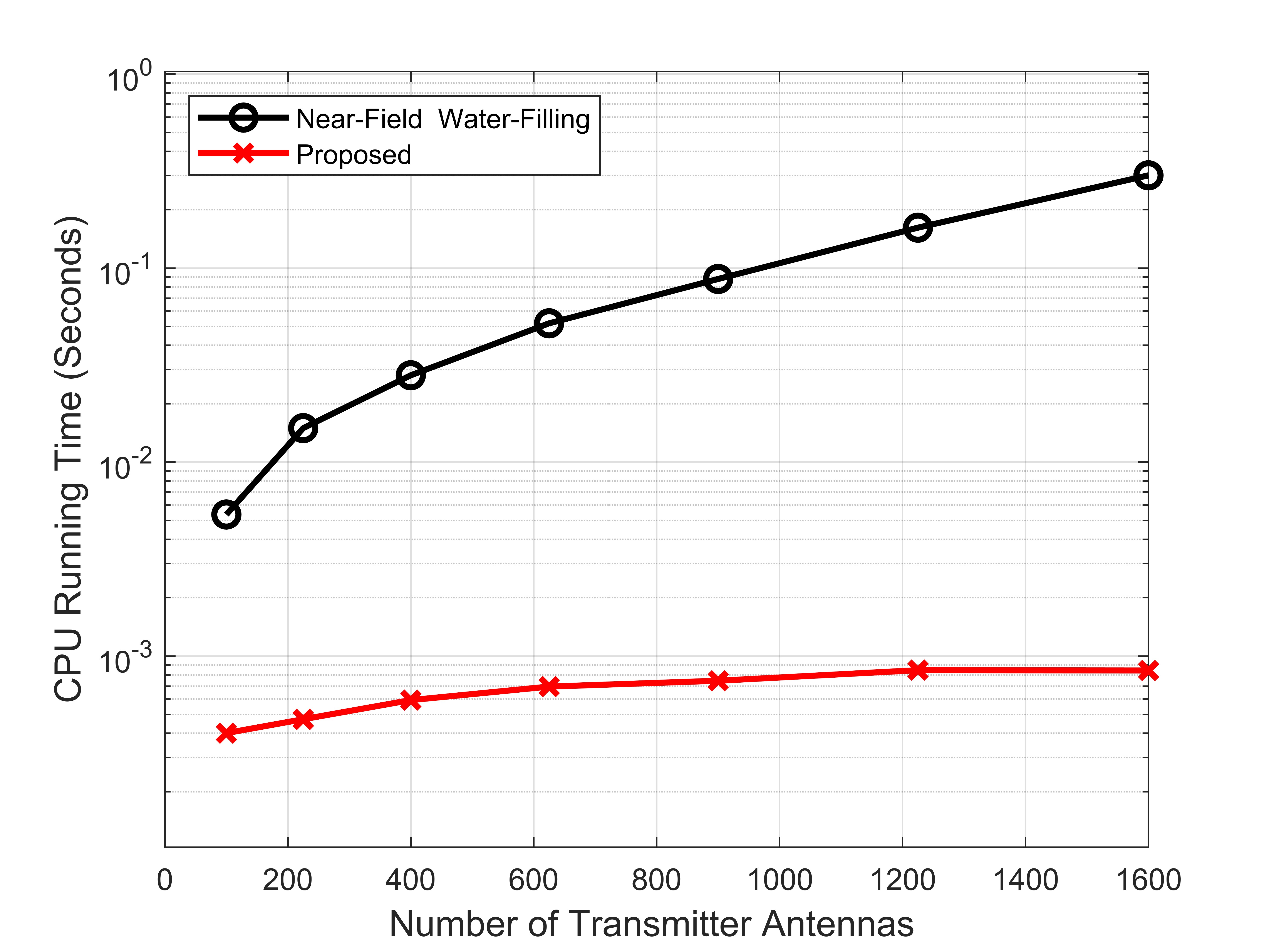}}
\caption{CPU running time versus the number of transmit antennas on an Intel Core i7-9750H CPU @ 2.60 GHz with 16 GB RAM.}
\label{CPU running}
\end{figure}

Fig. \ref{transmitter number_rate} and Fig. \ref{CPU running} evaluate the achievable rate and CPU running time versus the number of transmitter antennas, respectively. As shown in Fig. \ref{transmitter number_rate}, the proposed Kronecker-SVD beamforming design approaches the performance of the full-dimensional SVD with water-filling, while outperforming the equal-power and far-field SVD schemes. This indicates that the proposed geometric channel decomposition preserves the main structure of the original NF channel and that water-filling is necessary for efficient power allocation across different streams. Meanwhile, Fig. \ref{CPU running} shows that the proposed beamforming design requires much less CPU running time than the full-dimensional benchmark, since it replaces the SVD of the full channel matrix with two lower-dimensional SVDs. The running-time advantage becomes more evident as the number of transmitter antennas increases, demonstrating the scalability of the proposed design for large-scale planar arrays. Therefore, the proposed beamforming design achieves near-optimal rate performance with reduced computational cost.

\section{Conclusion}
\label{sec:Conclusion}

This paper investigated a near-field MIMO communication system equipped with dual UPAs. By developing a generalized geometric model to analyze the 3D distance between arbitrary antenna elements, we derived a closed-form  NF-FF boundary  for dual-UPA configurations. Furthermore, by exploiting the geometric structure of the arrays, the near-field channel matrix was decomposed into a Kronecker-product of two lower-dimensional matrices. This decomposition enabled a low-complexity beamforming design for achievable-rate maximization.  Numerical results validated that our derived boundary generalizes the conventional Rayleigh distance, and the proposed beamforming design achieves near-optimal rate performance with reduced computational complexity.

\begin{appendices}

\end{appendices}

\bibliographystyle{IEEEtran} 
\bibliography{references1}    

\vspace{12pt}

\end{document}